# CoVE: Towards Confidential Computing on RISC-V Platforms

Ravi Sahita
Rivos Inc.
Portland, Oregon, USA
ravi@rivosinc.com

Vedvyas Shanbhogue
Rivos Inc.
Austin, TX, USA
ved@rivosinc.com

Andrew Bresticker
Rivos Inc.
New York, NY, USA
abrestic@rivosinc.com

Atul Khare
Rivos Inc.
Seattle, WA, USA
atulkhare@rivosinc.com

Atish Patra
Rivos Inc.
Mtn. View, CA, USA
atishp@rivosinc.com

Samuel Ortiz
Rivos Inc.
Montpellier, France
sameo@rivosinc.com

Dylan Reid
Rivos Inc.
Mtn. View, CA, USA
dylan@rivosinc.com

Rajnesh Kanwal
Rivos Inc.
Watford, England, UK
rkanwal@rivosinc.com

## ABSTRACT

Multi-tenant computing platforms are typically comprised of several software and hardware components including platform firmware, host operating system kernel, virtualization monitor, and the actual tenant payloads that run on them (typically in a virtual machine, container, or application). This model is well established in large scale commercial deployment, but the downside is that all platform components and operators are in the Trusted Computing Base (TCB) of the tenant. This aspect is ill-suited for privacy-oriented workloads that aim to minimize the TCB footprint. Confidential computing presents a good stepping-stone towards providing a quantifiable TCB for computing. Confidential computing [1] requires the use of a HW-attested Trusted Execution Environments for data-in-use protection. The RISC-V architecture presents a strong foundation for meeting the requirements for Confidential Computing and other security paradigms in a clean slate manner. This paper describes a reference architecture and discusses ISA, non-ISA and system-on-chip (SoC) requirements for confidential computing on RISC-V Platforms. It discusses proposed ISA and non-ISA Extension for Confidential Virtual Machine for RISC-V platforms, referred to as CoVE.

## 1. RISC-V ISA and usages

A RISC-V hardware thread (hart) runs at a privilege level encoded as a mode in one or more CSRs (control and status registers). Three RISC-V privilege levels [2] currently defined are in Table 1. Privilege levels are used to provide protection between different components of the software stack and attempts to perform operations not permitted by the current privilege mode will cause an exception to be raised. These exceptions will normally cause traps into an underlying (higher privilege) execution environment.

| Level | Encoding | Name | Abbreviation |
|---|---|---|---|
| 0 | 00 | User/Application | U |
| 1 | 01 | Supervisor | S |
| 2 | 10 | Reserved | |
| 3 | 11 | Machine | M |

**Table 1: RISC-V privilege levels**

The machine level has the highest privileges and is the only mandatory privilege level for a RISC-V hardware platform. Code run in machine-mode (M-mode) is usually inherently trusted, as it has low-level access to the machine implementation. User-mode (U-mode) and supervisor-mode (S-mode) are intended for conventional application and operating system usage, respectively. Each privilege level has a core set of privileged ISA extensions with optional extensions and variants. For example, supervisor mode extended to support hypervisor execution and is utilized in CoVE (described in Section 3). Implementations might provide anywhere from 1 to 3 privilege modes trading off reduced isolation for lower implementation cost. All hardware implementations must provide M-mode, as this is the only mode that has unfettered access to the whole machine.

RISC-V supports privileged instructions to context switch in order to make requests to privileged environment or to return execution to a lower-privileged environment after trap-handling. ECALL is used to make a request to a higher privileged environment, and MRET and SRET are used to return to S-mode and U-mode privilege levels respectively.

## 2. Adversary and Basic Threat Model

The classes of **adversaries** considered are:
- Unprivileged Software adversary - includes software executing in User-mode managed by Supervisor-mode system software. This adversary can access user-mode CSRs, process/task memory, CPU registers in the process context.
- System Software adversary - includes system software executing in Supervisor or virtual-supervisor modes. Such an adversary can access privileged CSRs, assigned system memory, CPU registers, IO devices and induce device access.
- Startup Software adversary - includes system software executing at boot (Machine-mode), including BIOS, memory configuration, device firmware that can access system memory, CPU registers, critical IO devices and IOMMU.

- Simple Hardware adversary - includes hardware attacks such as bus interposers to snoop on memory/device interfaces, voltage/clock glitching, observe electromagnetic and other radiation, analyze power usage through instrumentation or tapping of power rails, which may give the adversary the ability to tamper with data in memory.
- Advanced Hardware adversary - includes attacks with unlimited physical access to the devices, and those that use mechanisms to tamper-with or reverse-engineer the hardware TCB, such as, key extraction using capabilities such as scanning electron microscopes, fib attacks, invasive de-cap, and supply chain attacks.
- Side/Covert Channel Adversary - includes attacks that leverage any explicit/implicit shared state (architectural or micro-architectural) to leak information across privilege boundaries via inference of characteristics from the shared resources (e.g., caches, branch prediction state, internal micro-architectural buffers, queues). Such attacks may require use of high-precision timers to leak information.

The main classes of **threats** to be considered are loss of confidentiality, integrity or replay-protection for a confidential workload's data-in-use. The data-in-use may be in memory or in internal hardware state and may be attacked via:
- in-band CPU access (software access via CPU load/store ISA).
- side-band CPU accesses due to software-induced side channels causing confidentiality loss which fall into these four sub-categories -
    - transient execution side-channel attacks in TCB components or workloads e.g., via shared caches, branch predictor poisoning, page-faults stepping.
    - architectural side-channel attacks due to shared cache and other shared resources e.g., prime/probe, flush/reload.
    - access to ciphertext with known plaintext to launch a dictionary attack on memory encryption keys.
    - side-channels due to performance monitoring and/or debug state.
- side-band software-induced memory changes e.g., row-hammer causing integrity loss.
- unauthorized external DMA from devices/accelerators.
- side-band CPU accesses by firmware by
    - physical memory aliasing
    - tampering with critical configurations (decoders, routing tables for IO fabrics)
- incorrect execution of workload, due to
    - malicious injection of faults.
    - dropping interrupts.
    - manipulation of time from M-mode CSRs.
- software or hardware attacks causing malicious address translation via attacks on MMU, IOMMU structures in memory, and associated translation caching structures in hardware (e.g., Translation lookaside buffers, paging level caches)

- hardware attacks on CPU/critical HW co-processors (e.g., RoT) conducted by
    - glitching, fault injection and other physical attacks.
    - exposed interface/links to other CPU sockets, memory.
    - hardware trojans, and supply chain attacks.
- cryptographic attacks by exploiting weakness in the cryptographic methods/protocols used for sealing confidential data or attestation.
- downgrading/forging TCB of the system (or TCB-derived information such as attestation evidence) to unsafe configurations, older (buggy) versions or loading/rolling back to malicious software, firmware components.

While protecting the confidentiality and integrity of tenant workload data-in-use is paramount; protecting against threats emanating from the workload on the host availability is also in scope –i.e., to prevent a confidential workload from causing a denial of service on the host platform. On the other hand, the reverse, protecting the availability of confidential workloads, is out of scope of this threat model.

## 3. RISC-V Hypervisor Extension

The RISC-V hypervisor extension virtualizes the supervisor-level architecture to support the efficient hosting of guest operating systems using a type-1 or type-2 hypervisor. The hypervisor extension changes supervisor mode into hypervisor-extended supervisor mode (HS-mode – See Table 2), where a hypervisor or a hosting-capable operating system runs. The hypervisor extension adds another stage (G-stage) of address translation, from guest physical addresses to supervisor physical addresses, to virtualize the memory and memory mapped I/O subsystems for a guest operating system. HS-mode acts the same as S-mode, but with additional instructions and CSRs that control the new stage of address translation and support hosting a guest OS in virtual S-mode (VS-mode). S-mode operating systems can execute without modification either in HS-mode or as VS-mode guests.

| Virtualization Mode (V) | Nominal Privilege | Abbreviation | Name | Two-Stage Translation |
|---|---|---|---|---|
| 0 | U | U-mode | User mode | Off |
| 0 | S | HS-mode | Hypervisor-extended supervisor mode | Off |
| 0 | M | M-mode | Machine mode | Off |
| 1 | U | VU-mode | Virtual user mode | On |
| 1 | S | VS-mode | Virtual supervisor mode | On |

**Table 2: RISC-V Privilege Levels and Virtualization Mode**

## 3.1 RISC-V CoVE

This section describes the proposed model of using the RISC-V Hypervisor extension to support CoVE. CoVE proposes non-ISA and ISA extensions for spatial and temporal isolation of a new HS-mode software module called the **Trusted Execution Environment Security Manager (TSM)** to manage security properties for workload assets to protect against access from the host operating system and hypervisor. Isolation of the TSM from the host is supported by ISA-extensions (described in Section 4) used by an HW-attested Machine-mode TCB component called the **TSM-driver**. The TSM enforces the isolation properties between workloads running as Confidential Virtual Machines using G-stage page tables. Confidential Virtual Machines are referred to as CoVE TEE Virtual Machines or **CoVE TVMs**.

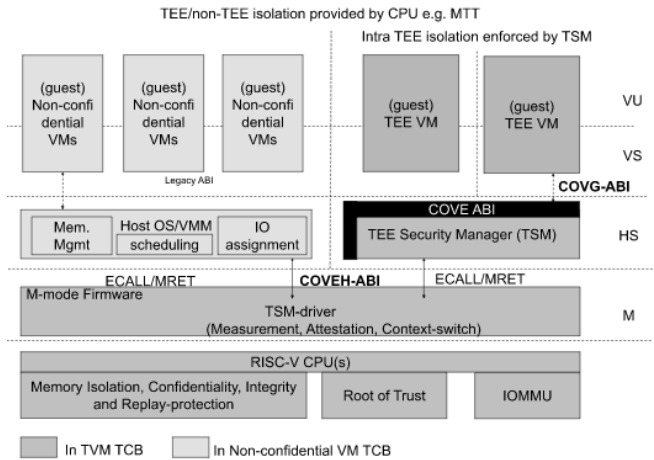

**Figure 1: RISC-V CoVE reference architecture**

The CoVE TCB consists of the TSM that acts as the TCB intermediary between TEE and non-TEE components. The SoC Root-of-trust (RoT) measures both the TSM-driver and the TSM to support attestation. The design of the TSM components is minimal in functionality – focusing only on security property enforcement for TEE VMs. The TCB also includes the hardware elements (processor and system-on-chip) that enforce confidentiality and integrity properties for workload data-in-use. The VMM is untrusted and continues to manage the resources for all workloads - confidential and non-confidential.

CoVE enables the OS/VMM to maintain the role of resource manager, with full bare-metal control of platform resources, even for the TVMs. The resources managed by the untrusted OS/VMM include memory, CPU, I/O resources and platform capabilities to execute the TVM workload. Using the H-extension as the ISA contract for confidential workloads allows most workloads to be moved to a TVM with no application re-factoring required and improves usability [4]. This approach enables the broadest use of ISA primitives and minimizes ISA changes specific to confidential workload reducing complexity, TCB and validation burden.

The TSM memory isolation from the host is achieved by new hardware isolation primitives proposed in CoVE called **Memory Tracking Table (MTT)** which allows the hypervisor to donate memory regions (and pages) to CoVE TVMs – with tracking and assignment policy set by the TSM-driver and the TSM. Memory isolated/access-controlled by the MTT is inaccessible to the host and may be additionally protected against physical access via cryptographic mechanisms (to enforce confidentiality, integrity and replay protection). Access to confidential memory regions is enforced in hardware via a **Confidential qualifier** maintained per hart. This mode/qualifier is enabled per-hart via TEECALL (a horizontal trap that is implemented via an ECALL and MRET), and disabled via TEERET – these context switch flows are supported by the TSM-driver. Access to confidential memory is allowed for the hart only when the confidential qualifier is set. Section 4 delves into the details of these extensions.

The TSM-driver delegates TSM isolation functions to the TSM, specifically, assignment and isolation of confidential memory to TVMs. The TSM-driver performs the following TCB functions:
- Bootstrap spatial and temporal isolation of the TSM via MMU extensions (MTT, IOMMU.)
- Context switching of hardware state for TSM execution (via ECALL, MRET)
- Firmware root-of-trust interface to the HW RoT to support attestation of the TCB.

The TSM interacts with the machine specific capabilities in the platform through the ABI it exposes to the host. The TSM provided ABI has the following functions:

- A **COVH-ABI** to manage the lifecycle of the TVM, such as creating TVMs, adding pages and virtual-harts to a TVM, scheduling a TVM virtual-hart for execution.
- An ABI to the TVM contexts, called **COVG-ABI,** to enable the TVM workload to request attestation functions, memory sharing/un-sharing functions or para-virtualized IO functions.
- A third **COVI-SBI** to manage secure interrupt facilities (using RISC-V Advanced Interrupt Architecture [11]).

The TSM functionality is by-design limited to support the necessary security primitives to ensure that the OS/VMM and non-confidential VMs do not violate the security of the TVMs through the resource management actions of the OS/VMM. These security primitives require the TSM to enforce TVM virtual-hart state save and restore, and invariants for memory assigned to the TVM (including G-stage translation). The host OS/VMM provides the typical VM resource management functionality for memory, IO. More than one TVM may be hosted by the host OS/VMM. Each TVM may consist of guest firmware, a guest OS and applications.

TVM execution is similar to a non-confidential VM operation with the exception that the TVM address space can be comprised of confidential and non-confidential regions. The former includes both measured pages (that are part of the initial TVM payload), and confidential zero-pages that can be mapped-in on demand by the VMM following runtime accesses by the TVM. The non-confidential TVM-defined regions include those for shared-pages used for para-virtualized IO. The TVM OS kernel is expected to be enlightened to use the COVG-ABI to retrieve attestation evidence to provide to a remote relying party to evaluate the CoVE TCB.

In Section 4 through Section 6, we describe the ISA, non-ISA (ABIs) and non-ISA (SoC) capabilities that form the building blocks of RISC-V CoVE.

## 4. ISA Capabilities

The following ISA extensions enforce the security properties for CoVE workloads and enable it to scale in a performant manner to large scale server platforms.

### 4.1 Confidential Qualifier for RISC-V hart

Isolating TVMs from the host software requires additional spatial and temporal isolation primitives. The goal is to allow host software to retain the privileged mode of operation while providing separation between non-TCB and TCB elements, specifically the host hypervisor, the TSM and TVM workloads. To address this gap, CoVE proposes a **"Confidential-mode qualifier"** maintained per processing element (e.g., hart) which is propagated into Physical Memory Protection (PMP), MMU lookups and the SoC fabric. This qualifier is minimally 1 bit of state in the hart; additional bits may be carried on the fabric to convey additional categories or modes of access to the platform. Table 3 shows the different modes of operation with the Confidential qualifier included. The PMP, MMU extensions for spatial isolation using this model are described in Section 4.2.

### 4.2. Domain Assignment of Memory and Devices

Beyond the Confidential qualifier on processing elements to enable TCB firmware to assert Confidential-mode of operation, we also require the ability to isolate physical address to enforce exclusive ownership of memory and devices to security domains (such as confidential domains) and map to properties such as confidentiality/integrity, even when the host software has full control of the existing access-control mechanisms such as page tables (MMU) and/or physical memory protection registers (PMP). This section describes the **Domain Assignment** Physical Memory Attribute (PMA) mechanism that addresses this requirement – this attribute is a dynamic property associated with memory that may be assigned or revoked for separate isolation domains. A domain in this usage is simply isolated physical address regions.

The physical memory map for a complete system includes various address ranges, some ranges may support specific properties, and some may not. In RISC-V systems, these properties of each region of the machine's physical address space are called physical memory attributes (PMAs). PMAs are checked by a hardware PMA Checker for any access to physical memory, including accesses that have undergone virtual to physical memory translation. Typically, all memory is equally accessible for software within a privilege level - there is no ISA mechanism to support isolation within a privilege level – this gap can be addressed by an additional level of dynamic PMA enforcement - a Domain Assignment PMA is proposed - the idea is to maintain meta-data per physical memory region (and page) that assigns domain ownership. The meta-data can be a minimal 1-bit (specified as C bit in this paper) or can hold additional meta-data and can be associated with PMPs on systems with only M or M/U support, or enforced via MMU extensions on systems that support M, S/H, and U levels. In this proposal, a MMU extension that looks up an in-memory **Memory Tracking Table (MTT Figure 2)** extension is used to enforce domain assignment.

**Table 3: Hart confidential qualifier and privilege modes**

| Virtualization Mode (V) | Nominal Privilege | Confidential Qualifier (C) | Abbreviation and Name | Address translation |
|---|---|---|---|---|
| 0 | U | 0 | U-mode (User mode) | [Single stage /bare] + [PMP] + **MTT** |
| 0 | S | 0 | HS-mode (Hypervisor-extended supervisor mode) | [Single stage/ /bare] + [PMP] + **MTT** |
| 0 | M | 0 | M-mode (Machine Mode) | Bare (TSM-driver in TCB) * |
| 1 | U | 0 | VU-mode (virtual user) | [Two-stage translation] + [PMP] + **MTT** |
| 1 | S | 0 | VS-mode (virtual supervisor) | |
| 1 | U | 1 | Confidential VU-mode (virtual user) | [Two-stage translation] + [PMP] + **MTT** |
| 1 | S | 1 | Confidential VS-mode (virtual supervisor) | |

*M-mode isolation is an ongoing discussion in the RISC-V security community (currently M-mode TCB may be restricted by de-privileging components to U or S-mode) – alternately the TSM-driver functionality may be subsumed into CPU ISA.

The hart with MTT enabled must enforce the following properties:
- A hart in non-Confidential-mode must not be able to access memory that is tracked via the MTT as confidential (isolated for a different isolation domain).
- A hart associated with a confidential workload must be the only ones allowed to access the memory owned by that workload per the MTT; such harts may choose to access globally shared memory to support IO use cases. Additional properties required for confidential workloads may be enforced, such as code fetch and page walks should always be via confidential memory.

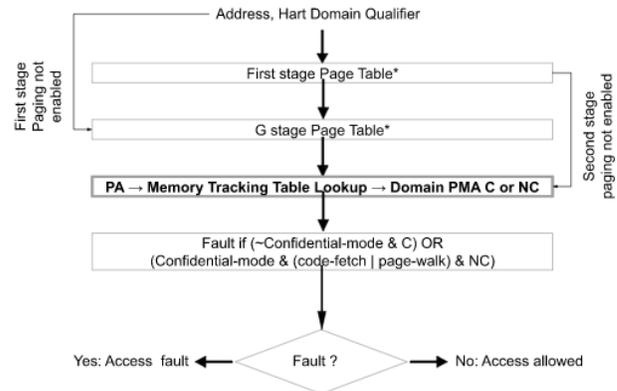

**Figure 2: Memory Tracking Table Enforcement**

The MTT capability remains under the control of the TCB – by associating its programming interface with the Confidential qualifier of the hart – thus allowing specifically the TSM-driver and the TSM (delegated explicitly by the TSM-driver) to dynamically manage it. The C bit (and potentially additional metadata bits can select cryptographic key(s) to enforce additional physical memory protection for confidential memory vs non-confidential memory. Implementations may choose to use separate keys associated with TVM workloads.

Another key requirement is the temporal isolation of the hart context state for the TSM – the memory isolation model described above can support the state isolation activated by a Domain context switch. In CoVE, these domain context switch flows may be implemented via non-ISA ABIs – the CoVE reference architecture (in Figure 1) describes new ECALL flows (TEECALL, TEERET) for isolation between the untrusted host hypervisor and the TSM – orchestrated via the TSM-driver in M-mode. This flow can also be subsumed by the hart and implemented as ISA to achieve the context switch using domain isolated memory.

The goal is to create a common flow to invoke TCB components which can be exercised via untrusted M-mode to invoke trusted M-mode components. The flow can be extended to lower privilege levels by the M-mode TCB component.

### 4.3. Confidential Interrupts

The RISC-V Advanced Interrupt Architecture [11] defines virtualization of interrupt files for VS-mode guests. The requirement introduced for confidential computing is to isolate (and reserve) guest interrupt files associated with TVMs (VS-mode level register files) – this is supported via the Confidential-mode qualifier for interrupt files, and via MTT for Memory-resident Interrupt Files. Access to the state of the TEE-assigned VS level register file when a hart is not in Confidential-mode must cause an illegal instruction exception (when V==0) or virtual instruction exception (when V==1). The interrupt delivery model is unchanged.

## 5. Non-ISA [SW ABIs]

The lifecycle of a CoVE TVM goes through phases of TVM build and initialization, execution and related runtime interactions with the host and finally teardown and reclamation of resources allocated to the TVM. These operations are carried out by the TSM when invoked by the host OS/VMM via the COVH-ABI or by the TVM via the COVG-ABI.

Executing confidential workloads in a CoVE TVM requires a sequence of one or more of the steps detailed below. These steps performed by the untrusted OS/VMM (host) or the TVM invoke the TSM to enforce security invariants.

- Platform TSM detection and capability enumeration – detects presence, version and capabilities of the TSM.
- Conversion of non-confidential memory to confidential memory: intrinsic that allows VMM to donate memory TVMs.
- Trusted VM (TVM) creation: intrinsic that allocates confidential memory to host TVM state.
- Donating confidential memory pages to the TSM for TVM page mapping via G-stage paging structures.
- Defining TVM memory regions: Allocates TVM address space as confidential or non-confidential and predicates subsequent memory assignment to these address spaces.
- Initializing TVM code and data payload to confidential-memory regions: as part of the TVM build process, this intrinsic drive the measurement of memory as well as zeroing non-measured memory pages.
- Creating TVM VCPUs: allocation of confidential memory pages to host TVM virtual hart context. The context structures are saved/restored by the TSM prior to/after TVM execution.
- Finalizing TVM creation: These intrinsic is used to complete the TVM measurement process for attestation.
- TVM execution: Intrinsic used for scheduling, injection of interrupts, and handling TVM faults and exits.
- Mapping TVM demand-zero confidential memory regions: enables lazy addition of memory into the TVM. The TSM verifies that assignment and mapping properties for confidential memory is not violated.
- Mapping TVM non-confidential shared pages – enables shared memory on-demand mapping into TVM shared memory regions used for IO or controlled CPU state sharing. The TSM verifies that assignment and mapping properties for confidential memory is not violated.
- Tearing down TVMs – intrinsic to stop a TVM execution and de-allocate confidential memory resources (back to the TSM).
- Reassignment of confidential memory – intrinsic to re-use unassigned confidential memory with the TSM enforcing mutual exclusive ownership of confidential memory to TVMs.
- Reclaiming confidential memory – intrinsic used to convert memory from confidential back to non-confidential so it can be used for non-confidential VMs/host workloads.

Details of the ABI functions can be found in the detailed RVI CoVE specification [3] developed in the AP-TEE Task Group.

## 6. Non-ISA [SoC Capabilities]

This section describes platform capabilities that supplement the hart capabilities required for confidential computing.

### 6.1 SoC Root-of-trust

The RoT hardware engine is the root of measurement, reporting and updates for the platform Trusted Compute Base (TCB). It manages the SoC security life cycle, key material and hence should be certified by the silicon manufacturer. The RoT verifies the configuration of SOC firmware and hardware modules and implements a Device Identity Composition Engine (TCG DICE [5]), where each TCB layer measures the next layer and provisions credentials. RoT-rooted measurements of workloads may then be verified by a relying party using an attestation protocol. An attestation protocol is used to verify the trust chain from hardware to the TCB firmware including the TSM-driver and the TSM. Each TVM derives its own attestation certificate from the TSM to support standard attestation protocols using a framework such as IETF RATS [6].

## 6.2. SoC considerations for Confidential Memory

Workload data-in-use stored outside the SoC package (such as memory) may require explicit cryptographic protection. Usage of memory on the platform starts off as untrusted resources in the non-confidential world and transition to their confidential analogue via the TSM which uses the SOC-specific mechanisms to activate protection for the memory against host accesses. Once the conversion process is complete, the VMM may assign confidential memory to a TVM via the TSM. A converted confidential resource can be freely assigned to another TVM when it is no longer in use. However, the VMM must reclaim an unused confidential resource for use in the non-confidential world (this is tracked and enforced by the TSM).

To meet these requirements, the SoC must map the domain assignment PMA to the transactions to SOC memory controller(s) and fabric to differentiate between confidential and non-confidential memory traffic. The isolation qualifiers (including Confidentiality-mode bit) must be enforced within the SOC caches using access-control, encryption, integrity and replay protection appropriate to the threat model being addressed. An example mapping is in Figure 3.

## 6.3. SoC Platform Data Protection

Large scale server platform requires Reliability, Availability and Serviceability (RAS). Also, platform debug and performance monitoring are critical functions to tune and debug workloads. However, these facilities can be leveraged as attack vectors and the SoC must provide appropriate controls to ensure that confidential data is not maliciously accessed via these mechanisms.

To mitigate attacks from platform mechanisms for RAS, QoS, debug and performance monitoring, appropriate authorization techniques [12] must be used to activate these mechanisms with opt-in controls provided to the TVM (recorded by the RoT and enforced by the TSM and TSM-driver) to give explicit control over TVM runtime state exposed to the host. Lastly, activation of debug and performance monitoring is explicitly reported in the attestation posture of specific TVMs.

## 6.4. SoC IO and Devices

Binding devices to TVMs expands the TCB of the TVM and brings in new threats, such as spoofing of device identifiers on the fabric, exposure of data on the link to the device, device firmware tamper, malicious programming of IOMMU and IO bridges.

Attaching devices to TVMs requires SOC support in I/O bridges and root ports to be able to protect the link from the IO bridge to the device. A pre-condition to set up end-to-end protection of the data sent to the device is to validate the authenticity of the device. There are active standardization activities for protocols to validate device certificates via DMTF SPDM [7], as well as protect the integrity and confidentiality of data on fabrics such as PCIe using IDE [8]. The TSM may control the trust state of device interfaces at a fine-granular assignment (virtual functions) via the PCIe TEE Device Interface Security Protocol (TDISP [9]). On the SOC side, RISC-V is pursing IO-PMP to be able to isolate IO controller programming interfaces to the TCB elements. Similarly, the proposed RISC-V IOMMU [10] can provide dedicated secure programming interfaces that can be enforced by the hart in Confidential-mode so that the TSM can program the IOMMU and enforce TVM security properties are met.

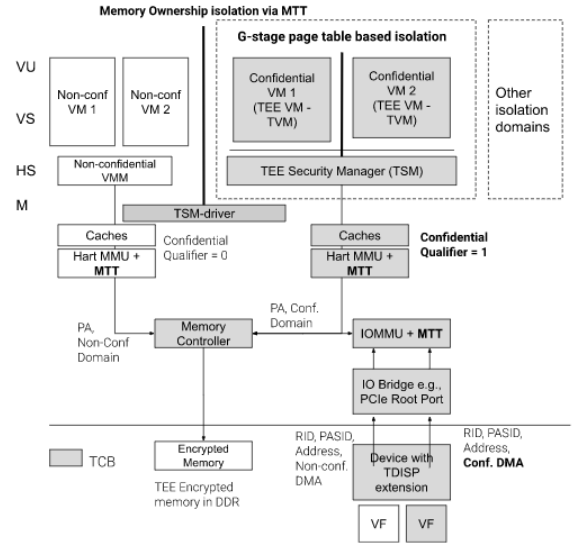

**Figure 3: SOC view of Conf. qualifier & domain assignment**

## 7. Related Work

This section is a brief survey of related commercial or academic TEE approaches.

Keystone [13] implements a security monitor using RISC-V M-mode firmware and uses PMPs to isolate enclaves. As defined, Keystone does not provide support for VM enclaves. Keystone enclaves are built out of contiguous memory which limits scalability for post-boot OS memory management for enclaves. Each enclave requires a PMP entry; the architecture can support N-2 enclaves for N hardware PMP registers. In contrast, CoVE uses the MTT to provide confidential memory isolation and uses the G-stage page tables for confidential memory assignment, simplifying OS memory management as it operates in architectural page sizes.

Intel TDX [14] executes TVMs in a mode called Secure Arbitration which is analog to functions of the CoVE TSM. The TDX module is a software component that operates in CPU VMX-root mode. Intel TDX splits the TVM address space into a private and shared address space and uses a Secure-EPT (equivalent of a G-stage page table in CoVE) to manage confidential memory assignment. Each Trust Domain is assigned a unique ephemeral key for memory encryption. TVM private memory and Secure-EPT pages are protected against tamper by using an optional cryptographic integrity mechanism using a truncated MAC store along with memory ECC (which restricts reliability mechanisms). Intel TDX supports remote attestation based on hardware-rooted keys. Intel TDX embeds the memory encryption key identifier in the physical address which restricts scalability.

AMD SEV-ES-SNP [15] uses a platform security processor (PSP) as the firmware TCB component to encrypt the TVM, with each TVM assigned an ephemeral memory encryption key. SNP uses a Reverse Map Table to track memory ownership by mapping a host address to the guest address associated with the page. However, the reverse lookup is enforced only for writes, allowing exposure to ciphertext. The PSP can limit scalability for large core systems and frequent trusted operations (such as page migration). AMD SEV provides support for hardware-rooted remote attestation.

ARM CCA [16] introduces Realm VMs that execute under the control of a Realm Management Monitor (RMM) – which executes as a peer to the untrusted hypervisor – it is the analog of the CoVE TSM. CCA extends ARM ISA to invoke the RMM from the host. Isolation between the realm and the other contexts is provided via a Granule Protection Table (GPT) used to track memory ownership on all accesses. Realms provides support for hardware remote attestation.

A common drawback of all the above approaches is the lack of support for Confidential IO – an ongoing area of work.

## 8. Conclusion and Future Work

The paper described CoVE, the first application-class processor framework for scalable support for Confidential VMs on RISC-V platforms running rich operating systems and hypervisors. The foundational extensions required for RISC-V platforms to support confidential VMs were described. These extensions are being developed in the RVI technical groups towards ratification on RISC-V platforms. An active open-source proof of concept implementation of CoVE is hosted at Github [17].

## ACKNOWLEDGMENTS

The authors would like to acknowledge the technical discussions and feedback from contributors in the RISC-V International forums where the ISA, non-ISA and platform specifications are being discussed for ratification, namely: Security Horizontal Committee (HC), Trusted Computing Special Interest Group (SIG), Application-Processor Trusted Execution Environment Task Group (TG), Privileged Architecture TG, Platform Runtime Services TG, Hypervisor SIG and SOC Infrastructure SIG.

## REFERENCES

[1] Confidential Computing Consortium (2022) Common Terminology for Confidential Computing, Available at: https://confidentialcomputing.io/wp-content/uploads/sites/85/2023/01/Common-Terminology-for-Confidential-Computing.pdf (Accessed: March 2, 2023).
[2] Waterman, A., Asanović, K., Hauser, J. et al. (2021) The RISC-V Instruction Set Manual Volume II: Privileged Architecture, GitHub. RISC-V International. Available at: https://github.com/riscv/riscv-isa-manual (Accessed: February 26, 2023).
[3] Sahita, R. et al. (2023) Confidential computing on RISC-V: Application-Processor Trusted Execution Environment (AP-TEE). GitHub. RISC-V International. Available at: https://github.com/riscv-non-isa/riscv-ap-tee (Accessed: February 26, 2023)
[4] Mark Russinovich, Manuel Costa, Cédric Fournet, David Chisnall, Antoine Delignat-Lavaud, Sylvan Clebsch, Kapil Vaswani, and Vikas Bhatia. (2021) Toward Confidential Cloud Computing: Extending hardware-enforced cryptographic protection to data while in use. Queue 19, 1, Pages 20 (January-February 2021), 28 pages. https://doi.org/10.1145/3454122.3456125
[5] Dominik Lorych and Lukas Jäger. (2022) Design Space Exploration of DICE. In The 17th International Conference on Availability, Reliability and Security
[6] H. Birkholz, D. Thaler, M. Richardson, N. Smith, and W. Pan. (2023) Remote ATtestation procedureS (RATS) Architecture. RFC Editor. Retrieved from https://www.rfc-editor.org/info/rfc9334 (Accessed: March 2, 2023)
[7] Brett Henning et al. (2022) Security Protocol and Data Mode (SPDM) Architecture White Paper. DMTF. DSP2058. Version 1.2.0. Available at: https://www.dmtf.org/sites/default/files/standards/documents/DSP2058_1.2.0.pdf (Accessed: March 2, 2023)
[8] Claire Ying. (2021) Why IDE Security Technology for PCIe and CXL? (2021). Cadence. Available at: https://www.chipestimate.com/Why-IDE-Security-Technology-for-PCIe-and-CXL/Cadence/blogs/3512 (Accessed: March 2, 2023)
[9] Richard Solomon. (2023) New PCIe TDISP Architecture Secures Device Interfaces with Virtual Servers. Synopsys. Available at: https://blogs.synopsys.com/from-silicon-to-software/2023/02/08/what-is-tdisp-pcie-io-virtualization/ (Accessed: March 2, 2023)
[10] Vedvyas Shanbhogue et al. (2023) RISC-V IOMMU Architecture Specification. Github. RISC-V International. Available at: https://github.com/riscv-non-isa/riscv-iommu/blob/main/riscv-iommu.pdf
[11] John Hauser et al. (2023) RISC-V Advanced Interrupt Architecture. Version 1.0-RC2. Github. RISC-V International. Available at: https://github.com/riscv/riscv-aia/releases/download/1.0-RC2/riscv-interrupts-1.0-RC2.pdf
[12] Tim Newsome et al. (2019) RISC-V External Debug Support version 0.13.2. RISC-V International. Available at: https://riscv.org/wp-content/uploads/2019/03/riscv-debug-release.pdf (Accessed: March 2, 2023)
[13] Dayeol Lee, David Kohlbrenner, Shweta Shinde, Krste Asanović, and Dawn Song. 2020. Keystone: an open framework for architecting trusted execution environments. In Proceedings of the Fifteenth European Conference on Computer Systems (EuroSys '20). Association for Computing Machinery, New York, NY, USA, Article 38, 1–16. https://doi.org/10.1145/3342195.3387532
[14] R. Sahita et al., 2021. Security analysis of confidential-compute instruction set architecture for virtualized workloads, 2021 International Symposium on Secure and Private Execution Environment Design (SEED), Washington, DC, USA, 2021, pp. 121-131, doi: 10.1109/SEED51797.2021.00024.
[15] Advanced Micro Devices, Inc. (2020), AMD SEV-SNP: Strengthening VM Isolation with Integrity Protection and More. AMD White Paper. Available at: https://www.amd.com/system/files/TechDocs/SEV-SNP-strengthening-vm-isolation-with-integrity-protection-and-more.pdf (Accessed: March 2, 2023)
[16] Mark Knight, Gareth Stockwell. October 2021. Arm Confidential Compute Architecture. In Proceedings of the Hardware and Architectural Support for Security and Privacy (HASP) 2021. Available at: https://haspworkshop.org/2021/slides/HASP-2021-Session2-Arm-CCA.pdf (Accessed: March 2, 2023)
[17] Salus: RISC-V hypervisor for TEE development (2023). GitHub. Rivos Inc. Available at: https://github.com/rivosinc/salus (Accessed: April 10, 2023)